\documentclass[12pt]{article}
\usepackage{graphicx}

\newsavebox{\sboxpubnumber}
\newsavebox{\sboxpubdate}
\newcommand{\pubdate}[1]{\begin{lrbox}{\sboxpubdate}{#1}\end{lrbox}}
\newcommand{\pubnumber}[1]{\begin{lrbox}{\sboxpubnumber}{\begin{tabular}{l} #1 \\
				 \usebox{\sboxpubdate}
				 \end{tabular}}
                           \end{lrbox}
                           \pubblock}
\newcommand{\Title}[1]{\begin{center} {\Large #1 } \end{center}}
\newcommand{\Author}[1]{\begin{center}{ \sc #1} \end{center}}
\newcommand{\Address}[1]{\begin{center}{ \it #1} \end{center}}
\newcommand{\andauth}{\begin{center}{and} \end{center}}
\newcommand{\pubblock}{\rightline{
			\usebox{\sboxpubnumber}}}
\newenvironment{Abstract}{\begin{quotation}  }{\end{quotation}}
\newenvironment{Presented}{\begin{quotation} \begin{center}
             PRESENTED BY P.~BUCCI AT\end{center}\bigskip
      \begin{center}\begin{large}}{\end{large}\end{center}
      \end{quotation}}

\def\beqra{\begin{eqnarray}}
\def\eeqra{\end{eqnarray}}
\def\beq{\begin{equation}}               
\def\eeq{\end{equation}}
\def\ds{\displaystyle}

\def                     \lta                     {\mathrel{\vcenter
     {\hbox{$<$}\nointerlineskip\hbox{$\sim$}}}}

     \renewcommand{\Re}{\mathop{\mathrm{Re}}}{\renewcommand{\Im}
     {\mathop{\mathrm{Im}}}
 \newcommand{\tr}{\mathop{\mathrm{tr}}}

 \def\e{e}
 \def\half{\mbox{\small
$\frac{1}{2}$}}  
\sloppy  
\input  epsf.sty     

\begin{document}

\begin{titlepage}
\pubdate{\today}                    
\pubnumber{XXX-XXXXX \\ YYY-YYYYYY} 

\vfill
\Title{Are Heavy Particles Boltzmann Suppressed?}
\vfill
\Author{Patrizia Bucci}
\Address{Department of Physics, University of Padova, \\
        Via F.  Marzolo  8, I-35131 Padova, Italy.\\
Institute of Theoretical Physics,\\
University of Warsaw,\\
ul. Hoza 69,PL-00-681, Warsaw, Poland. }
\vfill
\andauth
\vfill
\Author{Massimo Pietroni}
\Address{INFN - Sezione di Padova,\\
 Via F. Marzolo 8,
I-35131   Padova,  Italy}
\vfill
\begin{Abstract}
{Matsumoto and Yoshimura have recently argued that the number density
of heavy particles in a thermal bath is not necessarily Boltzmann-suppressed
for $T\ll M$, as power law corrections may emerge at higher orders in 
perturbation theory. This fact might have important implications on the
determination of WIMP relic densities.
On the other hand, the definition of number densities in a interacting 
theory is not a straightforward procedure.  
It usually requires renormalization of composite operators
and  operator mixing, which obscure the physical interpretation
of the computed thermal average.
We propose a new definition for the thermal average of a composite operator, 
which does not require any new renormalization counterterm and is thus
free from such ambiguities. Applying this definition to the annihilation
model of
Matsumoto and Yoshimura we find that it gives 
number densities which are Boltzmann-suppressed at any order in perturbation 
theory. We discuss also heavy particles which are unstable
 already at $T=0$, showing that power law corrections do in general emerge 
in this case. }
\end{Abstract}
\vfill
\begin{Presented}
    COSMO-01 \\
    Rovaniemi, Finland, \\
    August 29 -- September 4, 2001
\end{Presented}
\vfill
\end{titlepage}
\def\thefootnote{\fnsymbol{footnote}}
\setcounter{footnote}{0}

\section{Introduction}
The computation of number density of cosmological relics, such as weakly interacting 
heavy particles (WIMPs), neutrinos etc is usually realized by using a thermally average 
Boltzmann equation \cite{LW,KT}. The scenario which emerges is that of a sudden freeze-out: 
the particle number density follows the equilibrium value until the freeze-out temperature 
below which the annihilation is frozen. The freeze-out temperature of pair annihilation model 
may roughly be estimated by equating the annihilation rate to the Hubble rate, the relic 
abundance is then computed by the thermal number density at that freeze-out temperature, 
thus, we expect it is Boltzmann suppressed if the freeze-out temperature ($T_{f}$) is much smaller 
than the mass of the particle. This is the typical case for a WIMP: the freeze-out temperature 
is typically 4-5 $\%$ of the mass of the particle namely, they are non relativistic at 
decoupling, and their annihilation cross section is of the right order to give a contribution to $\Omega_{M}$ of order unity, so that they are very good candidates for cold dark matter models \cite{K}.\\
But from recent papers \cite{MY}, Matsumoto and  Yoshimura (MY)  have challenged the  
above conventional
conclusion, claiming  that two-loop corrections to  the number density
exhibit only  power-law suppression  in $T_f/M$, thus  dominating over
the  Boltzmann-suppressed  tree-level  contribution  if $T_f$  is  low
enough.  If confirmed,  such a  finding would  imply that  the present
constraints  on dark  matter  models are  actually underestimated  and
should be carefully  reconsidered. To be definite,  MY introduced a
toy model of two real scalar fields with Lagrangian 
\begin{eqnarray}
 {\cal L} &  = & \half \partial_\mu \phi \partial^\mu  \phi - \half M^2
\phi^2 -
 \frac{\lambda_{\phi}}{4!} \phi^4 +\half  \partial_\mu \chi
\partial^\mu  \chi  -
  \half  m^2   \chi^2  \nonumber  \\
  &  &  -
\frac{\lambda_{\chi}}{4!}       \chi^4-\frac{\lambda}{4}      \phi^2
\chi^2\,,\label{model}
\end{eqnarray}
with $M  \gg m$.  The following relations  were assumed among the
coupling           constants\[
           |\lambda_\phi|          \ll
\lambda^2\,,\;\;\;\;\;\;|\lambda|\ll|\lambda_\chi| < 1\,,
\]
so that the  light $\chi$ particles act as an  efficient heat bath for
the heavy $\phi$'s. They  then considered the quadratic  part of
the Hamiltonian for $\phi$,  
 \beq 
 H_\phi = \half \dot{\phi}^2
+ 
  \half (\nabla  \phi)^2 +\half M^2  \phi^2\,,
 \eeq 
  and, for
temperatures $T$ such  that $m \ll T \ll M$  they defined the number
density of  $\phi$ particles as 
  
\beq
 N_\phi=\frac{\langle H_\phi
+{\mathrm counterterms}\rangle_T}{M}\,\
\label{Ynumber}
\eeq 
where MY define the  thermal average for an operator $A$ as 
\beq \langle A \rangle_T \equiv  \frac{\tr A e^{- H/T}}{\tr e^{- H/T}}
- 
                \frac{\langle                0|A|0\rangle}{\langle
0|0\rangle},\label{Yaverage}
    \eeq
$H$    being    the   total Hamiltonian.   At   tree-level,   the   well  
known   Boltzmann-suppressed
contribution is obtained
\[ N_\phi^{(0)} = \left(\frac{M T}{2 \pi}\right)^{3/2} \e^{-M/T}\,.
\]
The power-law contribution arises at  two-loops and is given by \beq
N_\phi^{(2)}=                        c                       \lambda^2
\frac{T^6}{M^3},\,\,\,\,\,\,\,\,\,c=\frac{1}{69120}\,
\label{0}
 \eeq
While in the standard case the contribution to $\Omega_{M}$ depends on the mass 
of the particle only trough logarithmic terms, in this case, with the different power law behavior (\ref{0}), a simple calculation 
shows that the energy density goes like $\sim M^{1/4}$, implying much stronger constraints on the mass of the particle. 
The interpretation of (\ref{Ynumber}) as the physical
number density  of heavy $\phi$ particles has  been questioned 
by several authors 
\cite{SS,S,BJ}: what emerges from all these papers is that the real problem is 
 the definition 
of the number density for a heavy particle in a interacting theory.
We will show that it is possible to give a  
non-ambiguous definition of the number density of an interacting particle
at {\it any} temperature
and at any  order in perturbation theory. In the case  of the MY model
of   eq.~(\ref{model}),  our   definition   exhibits  the   expected
Boltzmann-suppression at low temperature. 
We will find it very convenient  to work in the  real-time
formalism of finite temperature field theory (RTF) \cite{LvW}, which 
has the advantage of keeping $T=0$ and finite temperature effects
well separated. Indeed, as we will discuss, the ambiguities
 in the conventional 
definitions of the number density are mostly due to the 
$T=0$-renormalization procedure. Being able to single out  
$T=0$  ultraviolet  (UV) 
divergences turns then out to be a great advantage when one is interested in 
genuine thermal effects.
The tree-level  propagator for a real scalar particle in the RTF has the structure: 
 \beq
 {\bf D_0}
= {\bf  D_0^{T=0}} +  {\bf D_0^T}  = i{\bf P[}\Delta_0{\bf  ]} +  i 
(\Delta_0-\Delta_0^*) {\cal N}(|q_0|) {\bf B}\,,
\label{RTFprop}
\eeq
where $\Delta_0^{(*)} =  (q^2-M^2 + (-) i \varepsilon)^{-1}$ is
the   tree-level  Feynman  propagator,    ${\cal   N}$   the Bose-Einstein
(Fermi-Dirac)     distribution    function,     ${\cal    N}(x)=
(\exp(x/T)-(+)1)^{-1}$    and   the   two
matrices ${\bf P}$ and ${\bf B}$ are defined as
\begin{eqnarray}
{\bf P[}a(q){\bf ]} & = & \left[
\begin{array}{cc}
a(q) & (a-a^*)(q)   \theta(-q_0)\\
   &  \\   
(a-a^*)(q)   \theta(q_0) & -a^*(q)
\end{array}
\right]\,\,, \nonumber \\
 & & \nonumber \\
{\bf B} & = & \left[ 
\begin{array}{cc}
1&1\\
 1&1
\end{array}
\right] , \nonumber \\
\end{eqnarray}
with $\theta(x)$ the Heaviside's step function. 

 The  full propagator  has the  same structure  as the  tree-level one,
eq. (\ref{RTFprop}),  with $\Delta_0^{(*)}$ replaced  by its all-order
counterpart         \cite{LvW}
        \beq
        \Delta^{(*)}(q)
=\frac{1}{q^2-M^2-\Pi(q_0,|\vec{q}|)+(-)i\varepsilon}\,\,,
\label{fulldelta}
\eeq
$ \Pi(q_0,|\vec{q}|)$ being the (renormalized) full self-energy 
at finite  temperature.  In the following,  we will refer  to $i {\bf 
P}[\Delta_0]$ and  $i {\bf P}[\Delta]$  as the ``$T=0$" parts  of the 
tree-level  and full  propagators, and  to $  i (\Delta_0-\Delta_0^*)
 {\cal N}(|q_0|)  {\bf B}$ and  $ i (\Delta-\Delta^*)  {\cal N}(|q_0|)
 {\bf B}$  as the  ``thermal" parts.  It  is important to  notice that
 what  we call  the  ``$T=0$"  part of  the  full propagator  actually
 contains thermal  corrections in the  self-energy, so {\it it  is not
 the  full $T=0$ propagator}.\\ The report presented here is a summary of our recent work \cite{pm}.

\section{Renormalization of $\phi_0^2$}
 Definition of number density generally comes 
out from bilinear operators (see (\ref{Ynumber}) for MY model ), thus,
 we have to study composite operator renormalization. They arise when we consider
local product of  fields making up the operator.  New UV
 divergences
are  induced when  it is  inserted in  a Green  function,  which are
generally   not   canceled  by   the   Lagrangian  counterterms.   The
renormalization of
 these  divergences then requires the introduction
of new
  counterterms. Renormalized  operators can be  defined, which
are generically
  expressible as linear combinations of  all the bare
operators of equal or lower
 canonical dimensionality (for a thorough
discussion of  composite operator
 renormalization,  see for instance
\cite{C,B}).   
Before considering  the complete energy-momentum  tensor, the analysis
of the
 composite operator  $\phi_0^2$ (where the $0$-label indicates
the bare  fields,
 masses and  coupling constants) will be  useful in
illustrating  the   main  points
  of   renormalization  and  thermal
averaging. We start by considering the
 $T=0$  case. In  the  model  (\ref{model}) the
  renormalization  of
$\phi_0^2$  then  induces a  mixing  with  $\chi_0^2$,  and the
  two
renormalized operators, $[\phi^2]$ and  
 $[\chi^2]$ can be expressed
as
\beq
\left(
\begin{array}{l}
\ds[\phi^2] \\ 
\ds[\chi^2]
\end{array}
\right)  =\left(\begin{array}{cc}   1-\delta  Z_{\phi\phi}&\delta
Z_{\phi\chi}\\
 \delta Z_{\chi\phi}&1-\delta Z_{\chi\chi}
\end{array}
\right)
 \left(
\begin{array}{l}
\phi_0^2 \\ \chi_0^2
\end{array}
\right)\,.
  \eeq
 In  minimal-subtraction  scheme  (MS)  the  bare
parameters can 
 be expressed  in terms of the renormalized ones as
\beqra                            
 &&\lambda_{i,0}=\mu^{4-D}\lambda_i
\left(1+\sum_{k=1}^{\infty}\frac{a_k^i(\{\lambda_j\};D)}{(D-4)^k}\right)\,,\nonumber\\
&&M_0^2=M^2    \left(1     +\frac{\delta    M^2}{M^2}\right)=    M^2
\left(1+\frac{\mu^2}{M^2}\sum_{k=1}^{\infty}\frac{b_k^M(\{\lambda_j\};D)}{(D-4)^k}\right)\,,\nonumber\\
&&m_0^2=m^2    \left(1     +\frac{\delta    m^2}{m^2}\right)=    m^2
\left(1+\frac{\mu^2}{m^2}\sum_{k=1}^{\infty}\frac{b_k^m(\{\lambda_j\};D)}{(D-4)^k}\right)\,,
\label{MS}
\eeqra
  where $\{\lambda_i\}= \{\lambda_\phi,\lambda_\chi,\lambda\}$, $D$ is the 
space-time dimensionality,
and  the important
  point is  
 that  the $a_k^i$'s,  $b_k^M$'s and
$b_k^m$'s are mass, momentum,  and
 temperature independent. 
 Using
eqs. (\ref{MS}) it is then possible  to prove that, for any renormalized
green function 
 $G$ the following relation holds
$$
 M_0^2  \frac{\partial G}{\partial M_0^2}+
  m_0^2 \frac{\partial
G}{\partial  m_0^2} =  
 M^2  \frac{\partial G}{\partial  M^2}+
 m^2
\frac{\partial G}{\partial m^2}\,,
$$
 showing  that, being the  right hand side manifestly  finite, the
insertion of  the combination $M_0^2  \phi_0^2 + m_0^2  \chi_0^2$ does
not  induce  any UV  divergence.  The  $\delta Z_{\phi\phi}$,  $\delta
Z_{\chi\chi}$ renormalization constants can then be expressed in terms
of the mass counterterms as
 \beq
 \left\{
\begin{array}{c}
\ds \delta  Z_{\phi\phi}=\ds -\frac{\delta M^2}{M^2}  - \frac{m^2}{M^2}\delta
Z_{\chi\phi}\\ \ds   \delta   Z_{\chi\chi}=-\frac{\delta   m^2}{m^2}   -
\frac{M^2}{m^2}\delta Z_{\phi\chi}
\end{array}
\right.  \eeq  whereas   the  remaining  counterterms  
  $\delta
Z_{\phi\chi}$, $\delta Z_{\chi\phi}$,
 can be computed as 
\begin{eqnarray}
\delta Z_{\phi\chi}  & = &\ds  -\left.\frac{\langle \phi_0^2(x) \chi(y)
\chi(z)\rangle}{\langle                                     \chi_0^2(x)
\chi(y)\chi(z)\rangle}\right|_{\mathrm
     DIV}\,,\,\,\,\,\,\nonumber
\\
 & & \nonumber \\ 
 \delta Z_{\chi\phi}& = & \ds -\left.\frac{\langle
\chi_0^2(x)      \phi(y)      \phi(z)\rangle}{\langle      \phi_0^2(x)
\phi(y)\phi(z)\rangle}\right|_{\mathrm DIV}\,.\nonumber
\end{eqnarray}

\section{Thermal average without mixing}
\begin{figure}[h]
\leavevmode
\hspace{0.5cm} \epsfbox{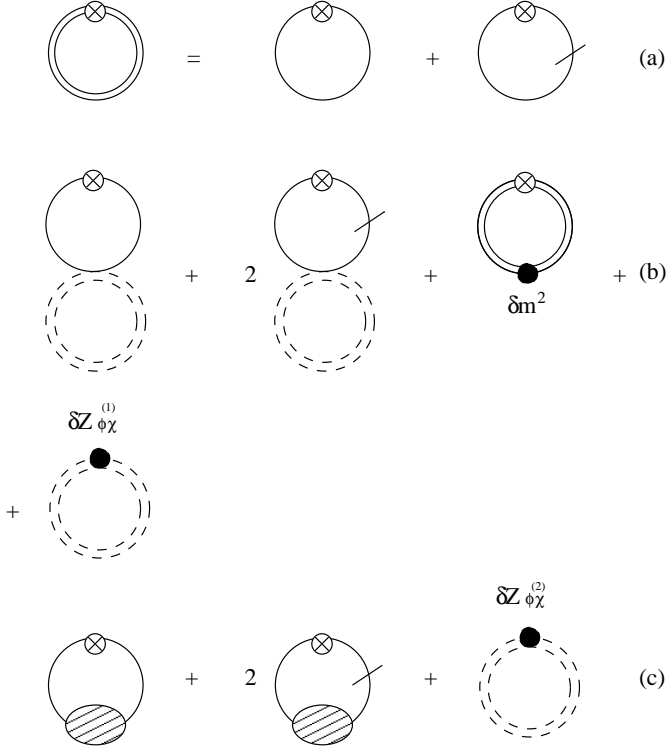}
\caption{Diagrammatic expansion of $\langle [\phi^2]\rangle_{T}$.} 
\label{fig1}
\end{figure}

The mixing between the $\phi^{2}$ and $\chi^{2}$ operator makes the definition of a pure $\phi$-number density quite cumbersome. 
A diagrammatic analysis can help us in identifying the $T=0$ nature of the pollution  by the $\chi$ field so that we can give a definition 
of thermal expectation value for $\phi^{2}$ which is free from it. 
Let us   consider  $ \tr   e^{-H/T}  [\phi^2]
/ \tr   e^{-H/T}$. 
At  tree-level in  the RTF it  is given by  the two  diagrams in
fig. \ref{fig1}a
\noindent
where the  continuous line  represents  the $T=0$  part of  the
$\phi$ propagator, the barred continuous  line the thermal part and the
double line  the sum  of the two.  The first diagram  is quadratically
divergent and  requires a subtraction, whereas the  second one, thanks
to  the  Bose-Einstein  function  contained in  ${\bf  D_0^T}(q)$,  is
finite. At  $O(\lambda)$ the diagrams of fig.~\ref{fig1}b contribute.  
There are also   contributions   $O(\lambda_\phi)$   
(such   as   higher   order
contributions    containing   also    $\lambda_{\chi}$)    but   their
consideration is  not necessary for  our discussion, so we  will limit
ourselves to contributions containing only powers of $\lambda$. 
 The (quadratic) divergence due to the $T=0$ $\chi$-tadpole diagram (dashed
line)  is  canceled  by  the  mass counterterm.
The  only  remaining
divergence comes from  the first diagram, in which both the
$\phi$  lines,  entering  the  cross  are  of $T=0$  type.  It  is  the
cancellation of this divergence which  calls into play the mixing, via
the  renormalization constant $\delta  Z_{\phi\chi}$ appearing  in the
last  diagram.
\vspace{1cm}
\\ At  $O(\lambda^2)$ (fig. \ref{fig1}c) the same  happens, where:
\begin{figure}[h]
\leavevmode
\hspace{2.5cm}
\includegraphics{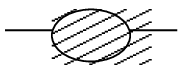}
\end{figure}
\noindent
\\
is the self energy to $O(\lambda^2)$ renormalized by Lagrangian 
counterterms.   
The  last
diagram, containing  a tree-level
 $\chi^2$ operator  multiplied by a
$O(\lambda^2)$  renormalization  constant is
  required  in order  to
cancel the divergence coming  from the integration over
 the momentum
of   $\phi$   particles   flowing    to   the   cross   with   $T=0$
propagators (the first diagram in fig. 1c).    
Note   that,   due    to   the    general   relation
$\sum_{i,j=1,2}\Sigma_{ij}=0$,  $\Sigma_{ij}$  being  the  self-energy  with
external
  thermal indices $i$  and $j$,  the contribution  with both
lines of thermal type
 also vanishes, once the summation over thermal
indices is  taken. The remaining
  loop divergences are  canceled by
usual Lagrangian counterterms and  therefore
 require no mixing. 
 The only divergence  which is not canceled by  neither Lagrangian nor
composite operators counterterms is the  tree-level one. Then, one usually
defines the  thermal expectation value of $\phi^2$  as the expectation
value of $[\phi^2]$  subtracted according to one of  the two following
ways  \cite{LvW,J},  
 \beqra
  \langle  [\phi^2] \rangle_T  &\equiv
&\frac{\tr [\phi^2]  
 e^{- H/T}}{\tr  e^{- H/T}} -  
 \frac{\langle
0|[\phi^2]|0\rangle}{\langle               0|0\rangle}              
\,\,\,\,\,\,\,\,\,\,\,\,\,\,\,{\mathrm     or}\nonumber\\
    &\equiv
&\frac{\tr [\phi^2] e^{- H/T}}{\tr e^{- H/T}} - \left.
 \frac{\langle
0|[\phi^2]|0\rangle}{\langle   0|0\rangle}
   \right|_{\mathrm   free
fields} \,\,\,\,\,\,\,\,\,\,\,\,\,\,\,.
\label{convthaverage}
\eeqra
 The definitions  above are both finite, and  thus a plausible
choice.  However,
 both of  them include  the contributions  from the
first and  the last graphs in 
figs.  \ref{fig1}b and \ref{fig1}c, that  is, they require mixing. 

The  above analysis  suggests a new  definition of thermal average  
of $\phi^2$,  which represents  the main  point of  this  paper, namely
\beqra
  \langle   \phi^2  \rangle_{\mathrm  Th}   &\equiv&  
  \int
\frac{d^4q}{(2\pi)^4}      D^T(q)\nonumber      \\     
      &=&\int
\frac{d^4q}{(2\pi)^4} 
 i(\Delta-\Delta^*)(q) {\cal N}(|q_0|)\,\,\,\,,
\label{thaverage}
\eeqra
 where  $ D^T(q)$  is any  of the four  component of  the {\it
thermal part of the full propagator}.
 
 Expanding $\Delta^{(*)}$ as
$\Delta^{(*)}=\Delta_0^{(*)}+\Delta_0^{(*)}   \Pi   \Delta_0^{(*)}   +
\dots$, and recalling the  relations between  $\Pi(q_0,|\vec{q}|)$ and
$\Sigma(q_0,|\vec{q}|)_{ij}$ \cite{LvW},
$$
  \Re  \Pi(q)  =  \Re \Sigma_{11}(q)\,,\,\,\,\,\,\,\,\,\Im  \Pi(q)  =
\varepsilon(q_0) \Im\left[\Sigma_{11}(q)+\Sigma_{12}(q)\right]\,,
$$
  one can  check that  the  perturbative expansion of fig.~\ref{fig1} is
reproduced, apart from the `problematic' diagrams. 
The definition  in (\ref{thaverage})  then automatically gets  rid of
all  the unpleasant  features of  
 the  conventional  definitions in
eq.   (\ref{convthaverage}),  namely,   the  need   of   an  arbitrary
subtraction and, more  importantly, composite operator renormalization
and mixing.  
\section{Averaging the energy-momentum tensor}
The  renormalization of  the  energy-momentum tensor  is discussed  in
refs.\cite{LvW,B,J}. Expressing  it in  terms of bare  parameters, the
operator
$$                                                        
T^{\mu\nu}=
\partial^\mu\phi_0\partial^\nu\phi_0+\partial^\mu\chi_0
\partial^\nu\chi_0-g^{\mu\nu}
{\cal L}_0+\cdots\,,
$$
is finite when inserted in a Green function, that is, it does not require
extra counterterms besides those already present in the Lagrangian.
The dots in the formula above 
  represent  pole   terms  proportional   to  $
(\partial^\mu \partial^\nu
 -  g^{\mu\nu} \partial^2)\phi_0^2$ and to
$ (\partial^\mu  \partial^\nu
 - g^{\mu\nu}  \partial^2)\chi_0^2$.
Since we are interested in the thermal average of $T^{\mu\nu}$
 alone,  translation invariance insures that such pole terms do not contribute.
On the other hand, a   subtraction    of    the   $T=0$    
contribution   as    in
eqs.  (\ref{convthaverage})  is   needed.  Working  with  renormalized
parameters,     $T^{\mu\nu}$      is     expressed     as
     \beq
T^{\mu\nu}=T^{\mu\nu}_{\mathrm can} +T^{\mu\nu}_{\mathrm c.t.}\,,
\label{tmunu} 
\eeq
 where $ T^{\mu\nu}_{\mathrm can}$  has the canonical form and $
T^{\mu\nu}_{\mathrm c.t.}$ contains  
{\it only the Lagrangian counterterms}.
 
The finiteness of $ T^{\mu\nu}$  means that the divergences induced by,
e.g.,  the  composite  operator  $\phi^2\chi^2$ are  canceled  by  the
Lagrangian  counterterms and the  particular combination  of operators
$\phi^4$,  $\chi^4$, $(\partial \phi)^2,\ldots$, appearing  in $
T^{\mu\nu}$. In general, if we split $ T^{\mu\nu}$ in parts, each part
separately  will require  composite operator  renormalization. Indeed,
this is  what is done  by MY in  ref. \cite{MY}. They split  the total
Hamiltonian as
$$
 T^{00}\equiv  H_{\mathrm tot} =  H_\phi+H_\chi+
 H_{\mathrm int}
+H_{\mathrm c.t.}\,,
$$
and compute thermal average of  $H_\phi$ (given in eq.~(2)).
In the RTF it gives
\begin{eqnarray}
\frac{\tr  H_\phi  e^{- H/T}}{\tr  e^{-  H/T}} &  =  &  
 \half  \int
\frac{d^4q}{(2\pi)^4}   \left[2q_0^2-    
   (q^2-M^2)\right]   \cdot
\nonumber \\  
 & & \nonumber \\
  & & \,\,\,\,\,\,\,\,\,\,\,\,\cdot
[{\bf D^{T=0}}+{\bf D^{T}}]_{12}(q)]\,,
\label{MYham}
\end{eqnarray}
where, differently from eq.  (\ref{thaverage}), also the $T=0$ part of
the full
 propagator appears. It  is important to recall here what we
have noticed after
 eq.(\ref{fulldelta}),  namely that the $T=0$ full
propagator  contains  thermal
  effects  via  the  self-energy.  Such
contributions  are not  canceled  by purely
  $T=0$ subtractions  like
those defined in  eq.  (\ref{convthaverage}).  By  perturbatively
  expanding
$[{\bf D^{T=0}}]_{12}$  in (\ref{MYham})  at $O(\lambda^2)$  we find,
apart from  divergent contributions  which require  composite
operator
  counterterms  to be  added  to
$H_\phi$, the ``famous" 
 power-law contribution
 \beq
\begin{array}{l}
\ds \half  \int  \frac{d^4q}{(2\pi)^4}  
 \left[2q_0^2-  (q^2-M^2)\right]
\left.[{\bf   D^{T=0}}]_{12}(q)\right|_{O(\lambda^2)}   
    =   \\
\ds \half  \int  \frac{d^4q}{(2\pi)^4}  
 \left[2q_0^2-  (q^2-M^2)\right] 
\theta(-q_0) \\
\ds
\left[ i(\Delta_0^2-{\Delta_0^*}^2) \Re \Pi-
(\Delta_0^2+{\Delta_0^*}^2) \epsilon(q_0) \Im \Pi\right] 
  \,   \\
\ds= {\mathrm div.} + \frac{1}{69120}\frac{\lambda^2 T^6}{M^2}+\cdots\,\,,
\end{array}
\label{powerlaw}
\eeq
where $\epsilon(q_0)\equiv \theta(q_0)-\theta(-q_0)$.
The  same contribution, with  opposite sign, is found  from the
corresponding  piece  of  the   thermal  average  for  $H_\chi$.
  
Paralleling our  discussion in the  previous section, we  will instead
define   the  energy  density   of  the   $\phi$  field   as
  
\beq
\rho_\phi^\prime\equiv   \half   \int   \frac{d^4q}{(2\pi)^4}   \left[2q_0^2-
(q^2-M^2)\right]
  [{\bf D^{T}}]_{12}(q)\,,
\label{newdef}
\eeq
  and analogously
for $\rho_\chi^\prime$. 
The  complete energy density will then  be split as
\beqra
 \rho_{\mathrm tot}\equiv
  \langle H_{\mathrm tot} \rangle_T
&\equiv &
 \frac{\tr  H_{\mathrm tot} e^{- H/T}}{\tr e^{-  H/T}} - 
\frac{\langle      0|      H_{\mathrm     tot}      |0\rangle}{\langle
0|0\rangle}\nonumber\\
   &&=\rho_\phi^\prime   +\rho_\chi^\prime  
+\rho_{\mathrm int}^\prime\,,
\label{split}
\eeqra
so that the contribution of eq.~(\ref{powerlaw}) enters 
$\rho_{\mathrm int}^\prime$ instead of $\rho_\phi^\prime$.
Notice that  neither
$\rho_{\phi}^\prime$ nor  $\rho_{\chi}^\prime$ are affected by the $T=0$ 
subtraction.

The definition in (\ref{split}) exhibits three remarkable properties:

{\it i)} The splitting in (\ref{split}) is closed under operator 
mixing and composite operator renormalization. No composite operator 
counterterm is required to make $\rho_{\phi}^\prime$ or  $\rho_{\chi}^\prime$
separately finite and, since $\rho_{\mathrm tot}$ 
is also finite, the same is true 
also for $\rho_{\mathrm int}^\prime$.
Thus, our definition of number 
density, differently from that considered by MY, is independent on the 
renormalization scale $\mu$ \footnote{Of course, at any finite order in 
perturbation theory we still have the usual $\mu$ dependence induced by
Lagrangian counterterms.}. There is no ``pollution'' from $\rho_{\chi}^\prime$
to $\rho_{\phi}^\prime$ at any value of $\mu$.

{\it ii)} None of the three pieces in (\ref{split}) contains 
$O(\lambda^2 T^6/M^2)$ terms. Indeed, the contribution of 
eq.~(\ref{powerlaw}) and  the analogous one for the $\chi$'s go both into 
$\rho_{int}^\prime$ and there, because of their opposite sign, cancel.

\begin{figure}[t]
\centerline{
\epsfbox{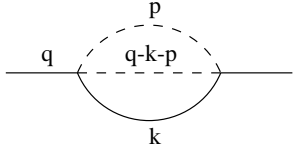}}
\caption{Two-loop contribution to the self energy .}
\label{figA1}
\end{figure}

{\it iii)} $\rho_\phi^\prime$ is Boltzmann-suppressed for $T\ll M$.Indeed, writing $[{\bf D^{T}}]_{12}(q)$ explicitly
\beq 
[{\bf D^{T}}]_{12}(q) = -
\frac{ 2 \Im \Pi(q) \epsilon(q_0)}{(q^2-M^2-\Re \Pi(q))^2 + \Im \Pi(q)^2} 
{\cal N}(|q_0|)\,,
\label{bbss}
\eeq
we see that, due to the Bose-Einstein function,  
non-Boltzmann-suppressed contributions to the integral in eq.~(\ref{newdef})
might only come from momenta $|q_0| \lta T \;(\ll M)$. But, for these values of
momenta, it is the 
imaginary part of the self-energy (Figure \ref{figA1}) which is Boltzmann-suppressed.
\noindent 
It can be easily understood looking 
at Figure 3 :
\begin{figure}[h]
\centerline
{\epsfbox{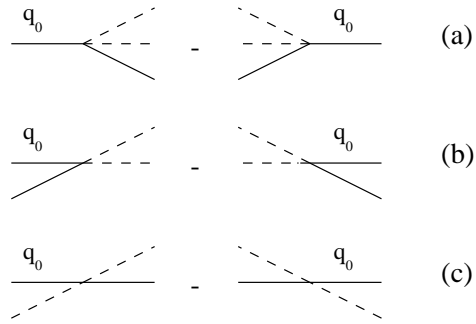}}
\caption{a) Decay, b) Annihilation, c) Landau damping
.}
\label{figA2}
\end{figure}
\noindent
\\for $|q_{0}|\lta T(\ll M)$ only the  annihilation 
 and the Landau damping  contribute. 
In the former case  (fig.~\ref{figA2}.b) 
the on-shell $\phi $ 
particle in the initial state comes from the heat-bath and then
carries a ${\cal N}$ factor. 
In the Landau damping case (fig.~\ref{figA2}.c), the energy to create
the on-shell $\phi$ has
to be provided by the $\chi$ from the
heat bath, so that a Boltzmann-suppressed ${\cal N}$ has  to be payed in 
this case too.

\section{A Better Definition for $N_\phi$}
Using the $\phi$-energy density as given in eq.~(\ref{newdef}), 
we can now define the $\phi$ number density for $T \ll M$ in a way 
analogous to MY's definition, eq.~(\ref{Ynumber}), that is
\[ 
N_\phi = \frac{\rho_\phi^\prime}{M}\;\;\;\;{\mathrm for\;\;\; T\ll M}\;.
\]
However a better definition, valid at {\it any} value
of $T$, and again based on the average of a quadratic operator, can be 
given as
\beq
N_\phi \equiv \int \frac{d^4 q}{(2\pi)^4} |q_0| [{\bf D^{T}}]_{12}(q)\,.
\label{verynew}
\eeq
As $\rho_\phi^\prime$, eq.~(\ref{verynew}) exhibits the nice features of
 not requiring composite operator renormalization and of being 
Boltzmann-suppressed at low temperatures, but in addition it may be extended 
to 
high temperatures or, equivalently, to the massless limit.
The origin of eq.~(\ref{verynew}) can be derived in analogy to the case of 
a complex scalar field, $\Phi$, in case an exact $U(1)$ symmetry is imposed to
the theory. The number density for the real scalar field $\phi$ should differ 
from the previous
quantity in two respects. First, particle and antiparticle contributions 
should be summed up, in order to account for their `Majorana' nature. This
is obtained by taking the modulus of $|q_0| $ inside (\ref{verynew}). 
Second, a factor $1/2$ is needed, in order to take into account  the 
number of degrees of freedom.
\section{The Decaying Case.}
The discussion of the previous sections is straightforwardly generalizable to
any model in which heavy particles annihilate into lighter ones, both bosons
and fermions. The basic conclusions remain unalterated, the main reason 
being the 
Boltzmann-suppression of the imaginary part of the self-energy at small 
momenta.

The situation changes drastically if the heavy particle is unstable  
at $T=0$. 
If, for definiteness,  it decays into two lighter particles of mass 
$m < M/2$ the propagator exhibits an 
imaginary part for $2 m < p_0 < M$ already at
 $T=0$, so that there is no reason to expect that it is Boltzmann-suppressed 
for $T<M$. The heavy particle is actually a resonance, whose energy-momentum 
can take values much lower than the peak at $p^2=M^2$, of course paying the 
usual Breit-Wigner suppression far away from the peak. In this case, power-law
 contributions to the resonance number density do in general emerge, as we will
briefly discuss.

We will consider for definiteness the annihilation model discussed in 
ref. \cite{decay} and in the first of refs. \cite{MY}, in which the interaction
  between the  heavy  $\Phi$ and the light $\chi$-bosons is due to the 
term
\beq
{\cal L}_{\mathrm int} = - \frac{\mu}{2}\Phi \chi^2\,,
\eeq
where we  assume  that $\mu \ll M$. In  this model, the 
imaginary part of the self-energy is $\propto \mu^2$, so that, using 
(\ref{bbss}) in (\ref{verynew}) we get 
\beq
N_\Phi \sim \frac{\Gamma_0}{M^3}T^5\,,
\eeq
with $\Gamma_0\equiv \mu^2/(32 \pi M)$ the $T=0$ decay rate on-shell.
In ref. \cite{decay} a different behavior, $N_\Phi \sim T^4$ was claimed.
Indeed, the following argument clarifies the physical origin of the power-law
contribution and confirms the $T^5$ behavior.

In the thermal bath, unstable $\Phi$ particles are continuously produced in 
$\chi$-$\chi$ annihilations. Being the $\Phi$ a resonance, the production 
energy, $\omega$, can be much smaller than $M$. 
The produced $\Phi$'s eventually decay with inverse lifetime 
$\Gamma=\mu^2/(32 \pi \omega) = \Gamma_0 M/\omega$.
Thus, the 
number density of $\Phi$'s of energy around $\omega$, $n^\Phi(\omega)$,
obeys the rate equation
\beq
\frac{d n^\Phi(\omega)}{dt}= \gamma - \Gamma n^\Phi(\omega)\,,
\eeq
where $\gamma$ is the rate per unit volume of the process
\beq
\chi\chi \rightarrow \Phi\rightarrow 
{\mathrm everything}\;.
\label{proc}
\eeq
The above equation leads to the equilibrium value
\[ 
n^\Phi_{\mathrm eq}(\omega) = \frac{\gamma}{\Gamma}\,.
\]
The production rate is given by $\gamma \sim N_\chi^2 \sigma v$, where the 
inclusive cross section $\sigma$  may be computed using the optical theorem as
$\sigma \sim \mu^2 \Im \Delta_\Phi / s$, $s$ being the square of the $\Phi$
four-momentum, and $\Delta_\Phi$ the full $\Phi$ propagator.

Taking $m\ll\sqrt{s} \sim T\ll M$, the initial $\chi$ particles are 
relativistic 
({\it i.e.} $N_\chi \sim T^3$) and we get $\sigma \sim \mu^4/(M^4 T^2)$.
Putting all together, we get a contribution to the total 
$\Phi$-number density from 
the region $\omega \sim T$ of order
\beq
n^\Phi_{\mathrm eq}(\omega\sim T)
 \sim \frac{\mu^2}{M^4}T^5 \sim \frac{\Gamma_0}{M^3} T^5\,.
\eeq
The $\Phi$'s with energies closer to the peak of the resonance 
live longer  but their contribution to the
number density  is 
Boltzmann-suppressed, as two highly energetic $\chi$'s are required in the 
initial state to produce them.

As the energy of the relevant $\Phi$'s is 
typically $O(T)$, the energy density is 
\[
\rho_\Phi \sim O\left(\frac{\Gamma_0}{M^3} T^6\right) + 
{\mathrm `Boltzmann-suppressed'}\,,
\]
and  $\rho_\Phi/\rho_\chi$ vanishes as $T^2$ at low temperatures.
Before concluding this section, we observe that the definition of a sensible
number density along the lines of eq.~(\ref{Ynumber}), {\it i.e.} starting 
from the free Hamiltonian, does not seem quite appropriate in this case. 
Indeed, as we have just discussed, the width of the $\Phi$ plays a crucial 
role, allowing energy values much lower than the tree-level mass-shell. It 
seems that, if one wants to insist in looking for definitions like 
eq.~(\ref{Ynumber}), a free Hamiltonian for quasi-particles, along the lines 
discussed for instance in ref.~\cite{Weldon}, should rather be used.


\section{Conclusions}
The definition of particle number density in a interacting theory is a delicate matter. The necessity of giving a meaning to divergent composite operator calls into play operator mixing, so that a clear separation between different particle species turns out to be a renormalization scale dependent procedure. We showed that in the RTF,  it is possible to give a proper definition of number density free from this problem. In the framework of MY annihilation model , our definition of the $\phi$-energy density is Boltzmann suppressed, no power suppressed terms of the type found by MY appear. This implies that the computations of
number density of  cosmological relics, on which dark matter models are based, are correct.\\
For an unstable particle , we showed, that power law contribution to the equilibrium number density do in general emerge and this may have important cosmological consequences for example in relation to the generation of the Baryon Asymmetry. It is clear that in this case the usual Boltzmann equation have to be modify to include off-shell effects. On this subject and the possible cosmological consequences of this new scenario, we are still working on \cite{P}.

\end{document}